\documentclass[final,5p,times,twocolumn,nopreprintline]{elsarticle}
\usepackage{amsmath,slashed,booktabs}
\usepackage{graphicx,graphics}
\usepackage{dcolumn}
\usepackage[hyperfootnotes=false]{hyperref}
\usepackage{xspace}
\usepackage{color}
\usepackage{balance}
\usepackage{multirow}
\usepackage{multicol}
\usepackage[normalem]{ulem}

\usepackage{fancyhdr}
\fancypagestyle{firstpage}{%
	
	\lhead{}
	\rhead{\small ACFI-T24-06, INT-PUB-24-045}
}
\usepackage{caption}
\pdfoutput=1 
\usepackage{graphicx}
\usepackage{dcolumn}
\usepackage{bm}
\usepackage{adjustbox}
\usepackage{multirow}
\usepackage{slashed}
\usepackage{color}
\usepackage[usenames, dvipsnames]{xcolor}
\usepackage[utf8]{inputenc}
\usepackage{tabularx,booktabs,subcaption}
\usepackage{hyperref}
\hypersetup{
    pdfnewwindow=true,   
    colorlinks=true,     
    linkcolor=blue,      
    citecolor=blue,      
    filecolor=blue,      
    urlcolor=blue        
}

\usepackage{marginnote}
\usepackage{cancel}
\usepackage{textcomp}
\usepackage{mathrsfs}
\usepackage{feynmp-auto}
\usepackage[normalem]{ulem}
\usepackage{empheq}
\usepackage{framed}
\usepackage{float}
\usepackage{bbm,dsfont}
\usepackage{setspace}
\usepackage{mathrsfs,enumitem,array}

\newcommand{\be}{\begin{equation}}
\newcommand{\ee}{\end{equation}}
\newcommand{\bea}{\begin{eqnarray}}
\newcommand{\eea}{\end{eqnarray}}
\newcommand{\lsim}{\buildrel < \over {_\sim}}

\newcommand{\onbb}{{0\nu\beta\beta}}
\newcommand{\TeV}{\,\mathrm{TeV}}
\newcommand{\GeV}{\,\mathrm{GeV}}
\newcommand{\MeV}{\,\mathrm{MeV}}
\newcommand{\eV}{\,\mathrm{eV}}

\newcommand{\fbi}{\,\text{fb}^{-1}}

\newcommand{\SU}[2]{SU(#1)_{\rm #2}}
\def\MWR{M_{W_R}}

\def\MDelta{M_{\delta_R^{\pm\pm}}}
\def\fRee{(f_R)_{ee}}
\def\invhalf{(T_{1/2}^{0\nu})^{-1}}
\definecolor{mydarkgreen}{RGB}{0, 100, 0} 

\def\nn{\nonumber}

\newcolumntype{C}{>{\centering\arraybackslash}p{2em}}
\newcolumntype{G}{>{\centering}p{2em}}

\usepackage{eso-pic}

\allowdisplaybreaks
\begin{document}

\renewcommand{\theequation}{\arabic{equation}}

\begin{frontmatter}
 
\title{
Dissecting Lepton Number Violating Interactions in the Left-Right Symmetric Model: $0\nu\beta\beta$ decay, Møller scattering, and collider searches
}

\author[SUNYATSET,SYSU]{Gang Li}
\author[UMass,TDLI,MOE,CALTECH]{Michael J. Ramsey-Musolf}
\author[UMass,Seattle,Evanston]{Sebasti\'an Urrutia Quiroga}
\author[SISSA]{Juan Carlos Vasquez}

\address[SUNYATSET]{School of Physics and Astronomy, Sun Yat-sen University, Zhuhai 519082, China}
\address[SYSU]{Guangdong Provincial Key Laboratory of Quantum Metrology and Sensing, Sun Yat-sen University, Zhuhai 519082, China}
\address[UMass]{Amherst Center for Fundamental Interactions, Department of Physics, University of Massachusetts, Amherst, MA 01003, USA}
\address[TDLI]{Tsung-Dao Lee Institute and School of Physics and Astronomy, Shanghai Jiao Tong University, 800 Dongchuan Road, Shanghai, 200240 China}
\address[MOE]{Shanghai Key Laboratory for Particle Physics and Cosmology, Key Laboratory for Particle Astrophysics and Cosmology (MOE), Shanghai Jiao Tong University, Shanghai 200240, China}
\address[CALTECH]{Kellogg Radiation Laboratory, California Institute of Technology, Pasadena, CA 91125, USA}
\address[Seattle]{Institute for Nuclear Theory, University of Washington, Seattle, WA 91195-1550, USA}
\address[Evanston]{Northwestern University, Department of Physics \& Astronomy, 2145 Sheridan Road, Evanston, IL 60208, USA}
\address[SISSA]{SISSA, Via Bonomea 265, I-34136 Trieste, Italy}

\begin{abstract}
In the context of the left-right symmetric model, we study the interplay of neutrinoless double beta ($0\nu\beta\beta$) decay, parity-violating Møller scattering, and high-energy colliders, resulting from the Yukawa interaction of the right-handed doubly-charged scalar to electrons, which could evade the severe constraints from charged lepton flavor violation. The $\onbb$ decay amplitude receives additional contributions from right-handed sterile neutrinos.  The half-life, calculated in the effective field theory framework, allows for an improved description of the contributions involving non-zero mixing between left- and right-handed $W$ bosons and those arising from exchanging a light right-handed neutrino.
We find that the relative sensitivities between the low-energy (or high-precision) and high-energy experiments are affected by the left-right mixing. On the other hand, our results show how the interplay of collider and low-energy searches provides a manner to explore regions that are inaccessible to $\onbb$ decay experiments.
\end{abstract}

\end{frontmatter}

\thispagestyle{firstpage}

\section{Introduction}
\label{sec:intro}
Explaining the origin of neutrino masses is a key open problem in particle and nuclear physics. If neutrinos are Majorana fermions, the mass term violates the lepton number by two units, $\Delta L =2$. The simplest tree-level realizations of the corresponding lepton-number-violating (LNV) Weinberg operator~\cite{Weinberg:1979sa} at dimension-5 are the type-I~\cite{Minkowski:1977sc, Gell-Mann:1979vob, Yanagida:1979as, Glashow:1979nm, Mohapatra:1979ia}, type-II~\cite{Konetschny:1977bn, Magg:1980ut, Schechter:1980gr, Mohapatra:1980yp, Lazarides:1980nt, Cheng:1980qt}, and type-III~\cite{Foot:1988aq, Ma:1998dn} seesaw mechanisms. Thus, in order to understand how neutrino masses are generated, it is essential to determine whether or not neutrinos are Majorana particles. The search for the neutrinoless double beta ($\onbb$) decay process  $(A, Z)\to (A, Z+2)+e^- + e^-$~\cite{Schechter:1981bd} provides one of the most powerful experimental probes of this question.\\

However, the detailed interpretation of $\onbb$ decay searches depends on the underlying mechanism. Beyond the \lq\lq standard mechanism\rq\rq\ involving the exchange of three light Majorana neutrinos, $\onbb$ decay may also receive non-standard contributions from other $\Delta L =2$ LNV sources beyond the Standard Model (BSM) if the associated mediator is sufficiently light. This possibility has been extensively studied in Refs.~\cite{Hirsch:1996qw, Faessler:1996ph, Prezeau:2003xn, Allanach:2009iv,  Tello:2010am, Bonnet:2012kh, Helo:2013dla, Helo:2013ika, Peng:2015haa, Cirigliano:2017djv, Cirigliano:2018yza, Cirigliano:2018yza, Li:2020flq, Harz:2021psp, Li:2021fvw,  Graesser:2022nkv, Li:2022cuq,Bolton:2021hje}. A systematic description of $\onbb$ decay can be performed in the effective field theory (EFT) approach, which provides an end-to-end framework from the LNV scale to the nuclear scale~\cite{Prezeau:2003xn, Cirigliano:2017djv, Cirigliano:2018yza}. \\

In order to distinguish different contributions to $\onbb$ decay, one needs to study the correlated signals in the high-energy and low-energy (or high-precision) frontiers~\cite{Cirigliano:2022oqy}. Complementary collider probes of LNV responsible for $\onbb$ decay at colliders have been explored in the context of ultraviolet (UV) complete models~\cite{Prezeau:2003xn,  Allanach:2009iv, Tello:2010am, Cirigliano:2018yza, Li:2020flq, Li:2022cuq,Bolton:2021hje} and simplified models~\cite{Bonnet:2012kh, Helo:2013dla, Helo:2013ika, Peng:2015haa, Harz:2021psp, Li:2021fvw, Graesser:2022nkv}. At the Large Hadron Collider (LHC), searching for same-sign charged leptons pairs plus di-jets~\cite{Keung:1983uu} provides a unique test of TeV-scale lepton number violation~\cite{Helo:2013ika,Peng:2015haa,Nemevsek:2018bbt, Harz:2021psp,Nemevsek:2023hwx}.
Besides the LNV searches, the lepton-number-conserving (LNC) searches, including direct searches for new particles and precision tests of SM interactions, offer a valuable diagnostic of the BSM
models, see for example Refs.~\cite{Helo:2013ika,Dev:2018sel,Du:2022vso}.\\

Working within the framework of the left-right symmetric model (LRSM)~\cite{Pati:1974yy,Mohapatra:1974gc,Mohapatra:1974hk,Senjanovic:1975rk,Senjanovic:1978ev}, we will study $\onbb$ decay and its interplay with precision measurements of parity-violating Møller scattering and collider searches, focusing on the doubly-charged scalar in the right-handed sector. The LRSM provides a natural origin of neutrino masses in the type-I and type-II seesaw mechanisms, in which right-handed neutrinos and scalar triplets are introduced. Due to the right-handed charged currents, non-standard contributions to $\onbb$ decay arise~\cite{Mohapatra:1980yp}, which may dominate over those from the exchange of light active neutrinos~\cite{Prezeau:2003xn, Tello:2010am, Li:2020flq}.
The scalar potential of the model is provided with doubly-charged scalars from the left-right scalar triplets. 
The mass and Yukawa couplings of the doubly-charged scalar in the left-handed sector are severely constrained by charged lepton flavor violation (CLFV)~\cite{Tello:2010am, Dinh:2012bp, Barry:2013xxa, Dev:2018sel, Barrie:2022ake}.
On the contrary, a doubly-charged scalar in the right-handed sector could have a mass much lower than its counterpart in the left-handed sector, as is the case in non-minimal LRSMs~\cite{Chang:1983fu,Sahu:2006pf,Deppisch:2014zta,Pritimita:2016fgr,Akhmedov:2024rvp} and its interactions can evade the CLFV constraints if the left-right symmetry is not imposed in the Yukawa interactions~\cite{Dev:2018sel,Akhmedov:2024rvp}\,\footnote{
In the minimal LRSM (mLRSM)~\cite{Mohapatra:1979ia,Mohapatra:1980yp}, the CLFV constraints on the right-handed doubly-charged scalar can also be avoided by assuming that its Yukawa coupling is diagonal~\cite{Fukuyama:2022dhe,Belfkir:2023lot}, and neutrinos obtain masses via the type-I seesaw mechanism.
}.
\\

The interaction of the doubly-charged scalar in the right-handed sector (hereafter right-handed doubly-charged scalar) with electrons not only induces a contribution to $\onbb$ decay but also to parity-violating asymmetries in polarized Møller scattering, Bhabha scattering, and the same-sign di-electron plus di-jet signature in proton-proton collisions. If Møller experiments or future Bhabha scattering in $e^+ e^-$ collisions observe deviations from the SM, such signals could be explained within non-manifest LRSM. To further determine whether the underlying physics is associated with lepton number violation, one can search for the $\onbb$ decay signals. Conversely, if a signal is observed in $\onbb$ decay, its origin in the LRSM can be tested by searching for deviations in Møller or Bhabha scattering experiments. In that case, compatibility with neutrino masses and CLFV constraints would again point toward some variant of the LRSM, possibly incorporating $D$-parity breaking~\cite{Chang:1983fu,Sahu:2006pf,Deppisch:2014zta,Akhmedov:2024rvp}.\\

The corresponding interplay among these observables
has been discussed in Ref.~\cite{Dev:2018sel}, where the analysis was focused on the regime where the doubly-charged scalar contribution dominated the $\onbb$ decay amplitude, and no left-right mixing was considered. Here, we will investigate the connections with the following new highlights:

\begin{itemize}
\item A possible non-zero $W_L-W_R$ mixing (also named left-right mixing) is considered. It has been shown in the chiral EFT framework~\cite{Prezeau:2003xn,Li:2020flq,Cirigliano:2018yza} that the contribution from the simultaneous exchange of $W_L$ and $W_R$ and right-handed neutrinos can dominate over the others to $\onbb$ decay because of a chiral enhancement at the hadronic level. 

\item The inclusion of all possible contributions to $\onbb$ decay in the LRSM. They correspond to the exchange of left-handed neutrinos, right-handed neutrinos, and doubly-charged scalars.

\item As a consequence of using the EFT description in $\onbb$ decay, we identify the parameter region where the right-handed neutrino transition between a light (dim-6) and a heavy (dim-9) EFT description exists, which proves to be essential for the sensitivity of $\onbb$-decay searches.
\end{itemize}

We obtain improved results in previously explored parameter regions and extend our exploration to the parameter space where the $\onbb$ decay amplitude is predominantly governed by light sterile neutrinos. This study also contributes to understanding the complementarity of $\onbb$ decay and low- and high-energy probes. On the one hand, we expand the relatively scarce literature connecting with PV Møller scattering, offering a more integrated treatment of these low-energy processes and providing important context for their interpretation.
On the other hand, we update existing collider constraints from dedicated studies \cite{Dev:2016dja, Nomura:2017abh, Dev:2018sel}, although that was not the exclusive focus of this work. We carefully explain the different collider constraints shown in our key figures, which now incorporate the very light sterile neutrino region and the small Yukawa coupling region. These areas have not been thoroughly explored in the literature before.\\

The remainder of the paper is arranged as follows. In Sec.~\ref{sec:LRSM}, we review the interactions, neutrino mixing, and left-right mixing in the LRSM. In Sec.~\ref{sec:exper_constr_high-E}, we discuss the current constraints and future prospects at high-energy colliders.
In Sec.~\ref{sec:exper_constr_low-E}, we discuss the current constraints and future prospects in low-energy experiments, including $\onbb$ decay and parity-violating Møller scattering.
In Sec.~\ref{sec:res_disc}, the results of $\onbb$ decay and combined sensitivities are presented and discussed in detail. We present our conclusions in Sec.~\ref{sec:conclusions}.

\section{Left-Right Symmetric Model}
\label{sec:LRSM}

The LRSM~\cite{Pati:1974yy,Mohapatra:1974gc,Mohapatra:1974hk,Senjanovic:1975rk,Senjanovic:1978ev}
extends the SM gauge group to $\SU3C\times \SU2L\times \SU2R \times  U(1)_{B-L}$, where $B$ and $L$ denote the SM abelian baryon and lepton quantum numbers. Henceforth, we denote the field representation under the $\SU2{R,L}$ and $U(1)_{B-L}$ groups by $(X_{\rm R},Y_{\rm L},Z_{B-L})$.
The scalar sector consists of one bidoublet $\Phi\in(2,2^\ast,0)$,
\begin{align}
\Phi = \begin{pmatrix}
\phi_1^0 & \phi_2^+\\
\phi_1^- & \phi_2^0 
\end{pmatrix}\,,
\end{align}
and two scalar triplets $\Delta_L\in (1,3,2)$, $\Delta_R\in(3,1,2)$
\begin{align}
\Delta_{L,R}=
\begin{pmatrix}
\delta_{L,R}^{+}/\sqrt{2} & \delta_{L,R}^{++}\\
\delta_{L,R}^{0}& -\delta_{L,R}^{+}/\sqrt{2}
\end{pmatrix}\,.
\end{align}

The extended gauge group is broken after the neutral scalars acquire vevs
\begin{align}
\langle \Delta_R\rangle &= 
\begin{pmatrix}
0 & 0 \\
v_R/\sqrt{2} & 0
\end{pmatrix}\;,\quad
\langle \Delta_L\rangle = 
\begin{pmatrix}
0 & 0 \\
v_L e^{{\rm i}\theta_L}/\sqrt{2} & 0
\end{pmatrix}\;,\nn\\
\langle \Phi \rangle &=
\begin{pmatrix}
\kappa/\sqrt{2} & 0\\
0 & \kappa^\prime e^{{\rm i}\alpha}/\sqrt{2}
\end{pmatrix}\;,
\end{align}
where all parameters are real~\cite{Deshpande:1990ip}. The right-handed scalar triplet vev, $v_R$, is taken to lie well above the electroweak scale $v = 246\GeV$. The bidoublet vevs, $\kappa$ and $\kappa'$, satisfy $\sqrt{\kappa^2 + \kappa^{\prime\,2}}=v$. The phases $\alpha$ and $\theta_L$ induce CP violation and will be set to zero ($\alpha,\,\theta_L=0$) for the purposes of this study. 
The scalar potential of the triplet and bidoublet sector follows the usual form \cite{Deshpande:1990ip,Dekens:2015csa,Maiezza:2016bzp,Maiezza:2016ybz}, leading to a spectrum that allows the doubly charged scalars to be relatively light~\cite{Akhmedov:2024rvp,Dev:2016dja}. This is because the squared mass, $M_{\delta_R^{\pm\pm}}^2 \simeq \rho_2 v_R^2$, is set by the dimensionless coupling $\rho_2$ from the scalar potential.
\\

Right-handed neutrinos appear in the doublet $L_R =(N,\ell_R)^{\sf T}$ and are thus charged under $\SU2R$ and couple to right-handed gauge bosons. It also couples to the scalar sector via the Yukawa interactions
\begin{align}
\label{eq:Lag-Yukawa}
\mathcal{L}_Y \ni& -\bar{Q}_L (\mathbf{\Gamma_q} \Phi + \mathbf{\widetilde{\Gamma}_q} \widetilde{\Phi}) Q_R -\bar{L}_L (\mathbf{\Gamma_l} \Phi + \mathbf{\widetilde{\Gamma}_l} \widetilde{\Phi}) L_R\nn\\
&-\bar L_L^c \,{\rm i}\tau_2\, \Delta_L \,\mathbf{f_L}\, L_L - \bar L_R^c \,{\rm i}\tau_2\, \Delta_R \,\mathbf{f_R}\, L_R + \text{H.c.}\,,
\end{align}
where $\widetilde{\Phi} \equiv \tau_2 \Phi^* \tau_2$ with $\tau_2$ being the second Pauli matrix, $\Psi^c_{L,R} = P_{R,L}\Psi^c$, and $\Psi^c = C \overline \Psi^{\sf T}$. The projectors $P_{R,L} = (1\pm \gamma_5)/2$ and $C$ is the usual charge conjugation matrix.

After the electroweak symmetry breaking (EWSB), the leptonic Yukawa interactions in Eq. \eqref{eq:Lag-Yukawa} give rise to neutrino masses through the type-I~\cite{Minkowski:1977sc, Gell-Mann:1979vob, Yanagida:1979as, Glashow:1979nm, Mohapatra:1979ia} and type-II~\cite{Konetschny:1977bn, Magg:1980ut, Schechter:1980gr, Mohapatra:1980yp, Lazarides:1980nt, Cheng:1980qt} seesaw mechanisms in terms of the symmetric $6\times 6$ matrix~\cite{Schechter:1981cv, Xing:2011zza}
\begin{align}
M_n = 
\begin{pmatrix}
M_L & M_D \\
M_D^{\sf T} & M_R
\end{pmatrix}\,,
\label{eq:Mn}
\end{align} 
where the Dirac and Majorana mass matrices are given by $M_D=(\kappa \mathbf{\Gamma_l} + \kappa^\prime \mathbf{\widetilde{\Gamma}_l})/\sqrt{2}$ and $M_{L,R}=\sqrt{2}\,\,\mathbf{f_{L,R}}\,v_{L,R}$, respectively,\footnote{If the left-right symmetry is not assumed in the Yukawa interactions, the phase $\theta_L$ contributes to $M_L$, but it does not affect the right-handed sector (unlike the scenario studied in Ref.~\cite{deVries:2022nyh}).
}. We diagonalize this neutrino mass matrix by using a $6\times6$ unitary matrix $U$,
\begin{equation}
m_\nu = \mathrm{diag}(m_1,\,m_2\dots ,\,m_6) = U^{\sf T} M_n^\dagger U\,,
\label{eq:mdiagU}
\end{equation}
and define the mass eigenstate $\widehat\nu = (\nu_1, \ldots, \nu_6)^{\sf T} \equiv \widehat\nu_m + \widehat\nu_m^c$ via
\begin{align}
\widehat\nu_m
= U^\dagger 
\begin{pmatrix}
\nu_L \\
N^{c}
\end{pmatrix}\,.
\end{align}

The neutrinos $\nu_{1,2,3}$ are active neutrinos, while $\nu_{4,5,6}$ are sterile neutrinos. We can decompose the unitary matrix $U$ as~\cite{Mitra:2011qr}
\begin{align}
U = \mathcal{U}_1 \mathcal{U}_2\,,
\end{align}
with 
\begin{align}
\mathcal{U}_1 = \begin{pmatrix}
1 & R \\
-R^\dagger & 1
\end{pmatrix}\;,\quad 
\mathcal{U}_2 = \begin{pmatrix}
U_{\text{PMNS}} & 0\\
0 & U_R
\end{pmatrix}\,,
\end{align}
where $U_{\mathrm{PMNS}}$ is the usual $3\times 3$ neutrino-mixing matrix, i.e. Pontecorvo-Maki-Nakagawa-Sakata (PMNS) matrix, $U_R$ is an additional $3\times3$ matrix and the matrix $R = M_D M_R^{-1}$ gives rise to the active-sterile mixing. This particular parametrization allows us to explicitly show the seesaw mechanism by noticing that~\cite{Mitra:2011qr}
\begin{align}
\mathcal{U}_1^\dagger M_n\,\mathcal{U}_1^\ast = \begin{pmatrix}
M_\nu & 0\\
0 & M_N
\end{pmatrix} \,,
\label{eq:block-diagonal-matrix}
\end{align}
where
\begin{align}
\label{eq:mneutrinos}
M_\nu &= M_L - M_D M_R^{-1} M_D^{\sf T}\,,\\
\label{eq:mneutrinos2}
M_N &= M_R\,.
\end{align}

The block-diagonal matrix is then diagonalized by the action of $\mathcal U_2$,
\begin{align}
\label{eq:diag2}
\mathcal{U}_2^\dagger \begin{pmatrix}
M_\nu & 0\\
0 & M_N
\end{pmatrix} \mathcal{U}_2^* = 
\begin{pmatrix}
\widehat{M}_\nu & 0\\
0 & \widehat{M}_N
\end{pmatrix}\;,
\end{align}
where the diagonal neutrino mass matrices are given by $\widehat{M}_\nu\equiv \mathrm{diag}(m_1,m_2,m_3)$ and $\widehat{M}_N \equiv \mathrm{diag}(m_4,m_5,m_6)$.\\

From Eq.~\eqref{eq:mneutrinos}, the type-I seesaw contribution to the mass matrix depends on both $M_D$ and $M_R=\sqrt{2}\, \mathbf{f_{R}}\, v_R$, while the type-II seesaw contribution is given by $M_L = \sqrt{2}\, \mathbf{f_{L}}\, v_L$. In the mLRSM, one can assume the type-I seesaw dominance, in which case ${\bf f_R}$ could be diagonal.
In the LRSM with spontaneous $D$-parity breaking~\cite{Chang:1983fu}, $v_L$ scales as $\sim\!v^2 v_R/M_{\slashed{P}}^2$, where $M_{\slashed{P}}$ denotes the $D$-parity symmetry breaking scale~\cite{Sahu:2006pf}. The active neutrinos can obtain masses either via the type-I seesaw or type-II seesaw mechanism~\cite{Sahu:2006pf,Deppisch:2014zta,Pritimita:2016fgr}.\\

For the purposes of our work, the charged electroweak gauge bosons are the relevant fields. After the breaking of the $\SU2{L,R}$ gauge symmetries, we define the mass eigenstates $W_{1,2}^{\pm}$ as
\begin{align}
\begin{pmatrix}
W_L^{\pm} \\
W_R^{\pm}
\end{pmatrix}
=\left(\begin{array}{rr}
\cos\zeta & -\sin\zeta\\
\sin\zeta & \cos\zeta
\end{array}\right)
\begin{pmatrix}
W_1^{\pm}\\
W_2^{\pm}
\end{pmatrix}\,,
\label{eq:MWmass_matrix}
\end{align}
where $\tan\zeta = {\kappa \kappa^\prime }/{v_R^2}$. Notice that, for an arbitrary CP-phase $\alpha$, one has $W_L^\pm = \cos\zeta\, W_1^\pm -\sin\zeta e^{\mp {\rm i}\alpha}\, W_2^\pm$~\cite{deVries:2022nyh}.
The charged gauge boson masses are 
\begin{align}
M_{W_1} &\simeq M_{W_L} = \dfrac{g_L v}{2}\equiv M_W\,,\nn\\
M_{W_2} &\simeq M_{W_R} = \dfrac{g_R v_R}{\sqrt{2}} \,,
\label{eq:MWmass}
\end{align}
where $g_{L,R}$ are the gauge couplings of $\SU2{L,R}$.
We introduce 
\begin{align}
\label{eq:lambda-xi}
\lambda = \dfrac{M_{W_1}^2}{M_{W_2}^2}\;,\quad
\tan\beta =\dfrac{\kappa^\prime}{\kappa}\equiv \xi\;,
\end{align}
such that the left-right mixing parameter $\zeta$ can be expressed as
\begin{eqnarray}
\tan\zeta = \lambda \sin(2\beta) = \frac{2\lambda\tan\beta}{1+\tan^2\beta} = \frac{2\lambda\xi}{1+\xi^2}\,.
\label{eq:tan_zeta__lambda_sin2beta}
\end{eqnarray}

Notice that in the mLRSM, $g_R \equiv g_L$, while the LRSM with $D$-parity breaking predicts the relation $g_R = 2 g_L/3$~\cite{Deppisch:2014zta}. 
In the context of an extended seesaw, however, there is no restriction on the relation between $g_R$ and $g_L$, as shown in Ref.~\cite{Pritimita:2016fgr}. For the purpose of our analysis, we will assume that $g_R = 2 g_L/3$. However, it is important to emphasize that our conclusions will remain valid regardless of the specific value chosen for $g_R$.

\section{Constraints and prospects at high-energy colliders}
\label{sec:exper_constr_high-E}

In this section, we study the current constraints and future sensitivities to the LRSM at high-energy colliders. Direct searches for the right-handed gauge boson $W_R$ at the LHC provide the most important diagnostic of the LRSM. The most stringent lower bound on the mass of $W_R$ is $6.4\TeV$ obtained by the ATLAS Collaboration~\cite{ATLAS:2023cjo} with the full Run-2 data of $138\fbi$.\\

The same-sign dilepton searches at the LHC have excluded the doubly-charged scalar with mass below $1080\GeV$ using its decays into $\ell^\pm \ell^\pm$ (with $\ell = e,\mu$)~\cite{ATLAS:2022pbd} with the full Run-2 data.
A future high-energy $pp$ collider like the Super Proton-Proton Collider (SPPC)~\cite{Tang:2015qga} or the Future Circular Collider (FCC-hh)~\cite{FCC:2018vvp}, operating at a center-of-mass energy of $100\TeV$, is expected to probe the doubly-charged scalar to higher masses. Pair-production could occur via the Drell-Yan process, and the reach for the doubly-charged scalar mass with an integrated luminosity of 30~ab$^{-1}$ could extend up to $5.1\TeV$ for $\MWR=7\TeV$ and $2.5\TeV$ for $\MWR\gtrsim 15\TeV$, respectively~\cite{Dev:2016dja}.\\

In addition to the direct searches, the Bhabha scattering process $e^+ e^- \to e^+ e^-$ places an indirect constraint on the mass of the doubly-charged scalar. Ref.~\cite{Nomura:2017abh} obtained $\MDelta/|\fRee| > 2.43\TeV$ based on the Large Electron-Positron (LEP) collider data~\cite{ALEPH:2013dgf}, improving the result derived in Ref. \cite{Dev:2018sel}. We follow the approach in Ref. \cite{Dev:2018sel} to set a conservative limit for the reach of future $e^+e^-$ colliders by rescaling the LEP limit by a factor of 
\begin{equation}
\sqrt{(\sigma_{\rm LEP}\mathcal L_{\rm LEP})/(\sigma_{e^+e^-}\mathcal L_{e^+e^-})}\,.\label{eq:factor}   
\end{equation}

For LEP, we consider an average center-of-mass energy of 195.6 GeV with an integrated luminosity of 745 pb$^{-1}$ \cite{ALEPH:2013dgf}.
We will focus on the Circular Electron Positron Collider (CEPC)~\cite{CEPCPhysicsStudyGroup:2022uwl} at 240 GeV with an integrated luminosity of 20 ab$^{-1}$, and the Future Circular Collider (FCC-ee)~\cite{Agapov:2022bhm} at 240 GeV with an integrated luminosity of 5.1 ab$^{-1}$.

\section{Constraints and prospects in low-energy experiments}
\label{sec:exper_constr_low-E}

In this section, we will investigate the sensitivities to the LRSM in the low-energy experiments. The presence of right-handed charged currents impacts the mixings of $K$ and $B$ mesons, which have been investigated in Refs.~\cite{Bertolini:2014sua, Dekens:2021bro, Zhang:2007da}. However, the resulting constraints on the mass of $W_R$ are weaker than those from direct searches at the LHC~\cite{Li:2020flq, ATLAS:2023cjo}. The experimental constraint on $\MWR$ in Sec. \ref{sec:exper_constr_high-E} can be directly translated into the equivalent bound $\lambda < 1.58 \times 10^{-4}$ using Eq.~\eqref{eq:lambda-xi}. Additionally, $\xi<0.8$ is required to ensure the validity of perturbative expansions in the LRSM scalar sector~\cite{Maiezza:2010ic, Nemevsek:2023yjl}.\\

The left-right mixing angle $\zeta$ is related to the parameters $\xi$ and $\lambda$ as shown in Eq.~\eqref{eq:tan_zeta__lambda_sin2beta}.
It is important to highlight that to resolve the deviation from the CKM unitarity using the determination of $|V_{ud}|$ from the nuclear superallowed $0^+ \to 0^+$ decay \cite{Seng:2018yzq, Seng:2018qru, Czarnecki:2019mwq}, the left-right mixing angle $\zeta \in [4.5,11]\times 10^{-4}$ is requisite~\cite{Dekens:2021bro}. This leads to an apparent incompatibility between the CKM and collider implications. However, for the scope of this analysis, we will refrain from delving into the resolution of the discrepancy, and we will adopt the collider bounds.\\

Besides, $\xi$ and $\lambda$ can also be simultaneously constrained by the $\rho$ parameter~\cite{Czakon:1999ga}
\begin{align}
\rho = 1 + \left[-\left(\frac{2\xi}{1+\xi^2}\right)^2 +  \frac{(1-\tan^2\theta_W)^2}{2}\right]\lambda\,,
\label{eq:rho_beta-lambda}
\end{align}
where $\theta_W$ is the weak mixing angle. Using the latest global fits of electroweak precision measurements on the $\rho$ parameter, we set a bound on the $W_L-W_R$ mixing angle $\zeta$ and $M_{W_R}$.\\ 

In our analysis, we chose different benchmark values consistent with the aforementioned constraints:
(I) $M_{W_R}=7\TeV$, $\xi=0.35$, (II) $M_{W_R}=15\TeV$, $\xi=0.75$, and (III) $M_{W_R}=25\TeV$, $\xi=0.8$. The benchmarks for $\xi=0$ and $M_{W_R}=7$, $15$ and $25\TeV$ are also considered.\\

As mentioned in Sec.~\ref{sec:intro}, searches for CLFV in $\mu\to e\gamma$, $\mu \to eee$, and $\mu\to e$ conversion in nuclei have set stringent constraints on 
physics beyond the standard model~\cite{Lindner:2016bgg}. 
The constraints on the left-handed doubly-charged scalar $\delta_L^{\pm\pm}$ are~\cite{Dev:2018sel,Akeroyd:2009nu}
\begin{equation}
    \begin{aligned}
    \label{eq:cLFV-constraints}
         &\mu\to eee:\ M_{\delta_L^{ \pm\pm}}\left| \left(f_L\right)_{e e}^{\dagger}\left(f_L\right)_{e \mu}\right|^{-1/2} > 208\TeV\;,\\
   &\mu\to e\gamma:\ M_{\delta_L^{ \pm\pm}}\left|\sum_{\ell}\left(f_L\right)_{\mu \ell}^{\dagger}\left(f_L\right)_{e \ell}\right|^{-1/2} > 65\TeV\;,
    \end{aligned}
\end{equation}
where $M_{\delta_L^{ \pm\pm}}$ denotes the mass of $\delta_L^{ \pm\pm}$.\\

The right-handed doubly-charged scalar $\delta_R^{ \pm+}$, however, could evade these constraints if the left-right symmetry is not assumed in the Yukawa sector. In the LRSM with $D$-parity breaking, this is evident since parity and the $SU(2)_{\rm R}$ breaking scale decouple from each other~\cite{Chang:1983fu}, which has been considered in Ref.~\cite{Dev:2018sel}. A realistic model~\cite{Akhmedov:2024rvp} was recently developed along this line, making it feasible to explore the scenario where $\mathbf{f_{R}} \neq \mathbf{f_{L}}$.\\

\subsection{Neutrinoless double beta decay}
\label{subsec:0nbb}

Although the literature contains numerous studies regarding $\onbb$ decay in the LRSM (see, for example, Refs \cite{Tello:2010am, Awasthi:2015ota, BhupalDev:2014qbx, Bambhaniya:2015ipg}), the present study includes several novel elements. In particular, results in many of the previous analyses rely upon an expansion of the neutrino propagator $1/(\langle p\rangle^2 + m_N^2)$ in powers of $\langle p\rangle/m_N$, where $\langle p\rangle\simeq100\,{\rm MeV}$ is the typical momentum transfer and $m_N$ is the right-handed neutrino mass. 
It has been demonstrated that the $\onbb$ decay rate calculated using this approach can benefit from the systematic prescription based on chiral EFT, leading to sizable differences, particularly in cases involving chiral enhancement arising from left-right mixing at the quark level; see, for instance, Refs. \cite{Cirigliano:2018yza,Li:2020flq,Dekens:2020ttz,Cirigliano:2024ccq,Cirigliano:2025vye}. 
Additionally, the propagator-expansion approach breaks down for light sterile neutrinos, namely $m_N \lesssim \langle p\rangle$. Our work is based on a chiral EFT approach~\cite{Dekens:2020ttz}, which allows us to investigate the effects of a wider range of sterile neutrino masses.\\

The relation in Eq.~\eqref{eq:mneutrinos2} indicates that the mass of the right-handed neutrino is correlated with the magnitude of the Yukawa coupling, $\widehat M_N \sim {\bf f_R}\,v_R$, where the indices are omitted for simplicity.
Exploring small couplings would imply light right-handed neutrinos. If $m_{4,5,6}\lsim \Lambda_\chi$,
the right-handed neutrino fields should be kept as dynamical degrees of freedom in the approach of neutrino-extended SMEFT ($\nu$SMEFT). We adopt the framework developed in Refs.~\cite{Dekens:2020ttz, deVries:2022nyh} to study the contributions to the half-life of $\onbb$ decay from right-handed neutrinos with arbitrary mass.\\

\begin{figure}
\captionsetup[subfigure]{justification=centering}
     \centering
     \begin{subfigure}[b]{0.22\textwidth}
         \centering
         \includegraphics[width=\textwidth]{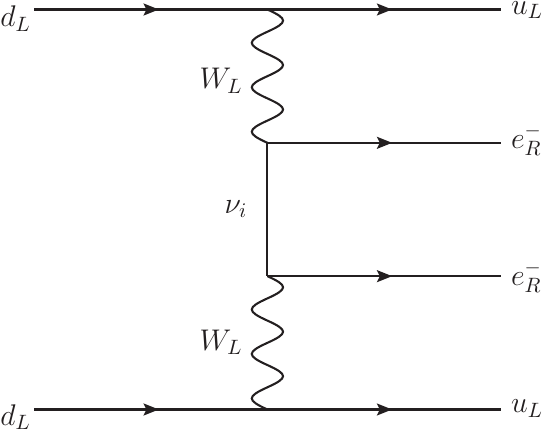}
         \caption{ }
     \end{subfigure}
     \hfill
     \begin{subfigure}[b]{0.22\textwidth}
         \centering
         \includegraphics[width=\textwidth]{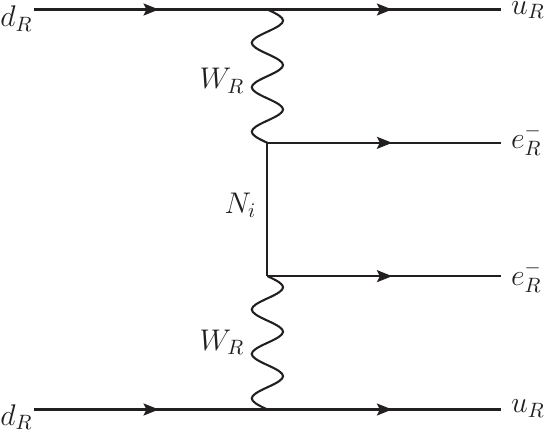}
         \caption{ }
     \end{subfigure}
     \hfill\par\bigskip
     \begin{subfigure}[b]{0.22\textwidth}
         \centering
         \includegraphics[width=\textwidth]{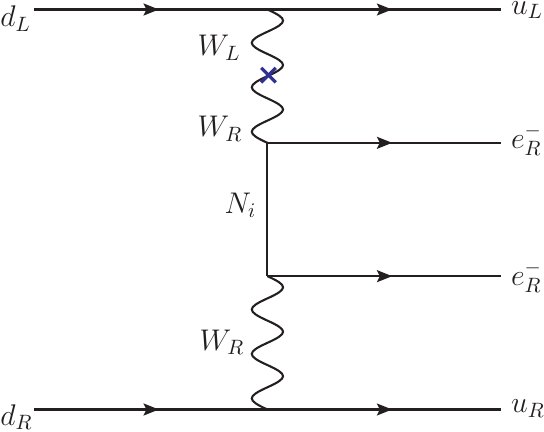}
         \caption{ }
     \end{subfigure}
     \hfill
     \begin{subfigure}[b]{0.22\textwidth}
         \centering
         \includegraphics[width=\textwidth]{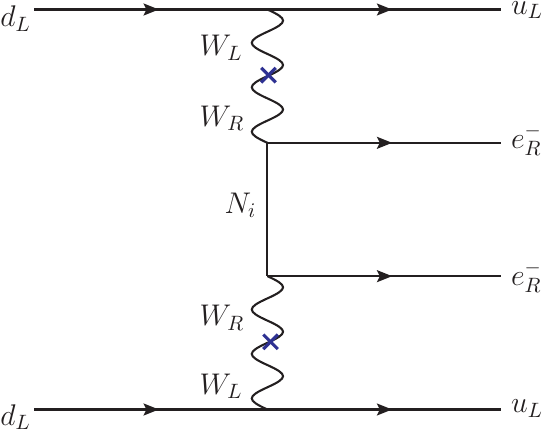}
         \caption{ }
     \end{subfigure}
     \hfill\par\bigskip
     \begin{subfigure}[b]{0.22\textwidth}
         \centering
         \includegraphics[width=\textwidth]{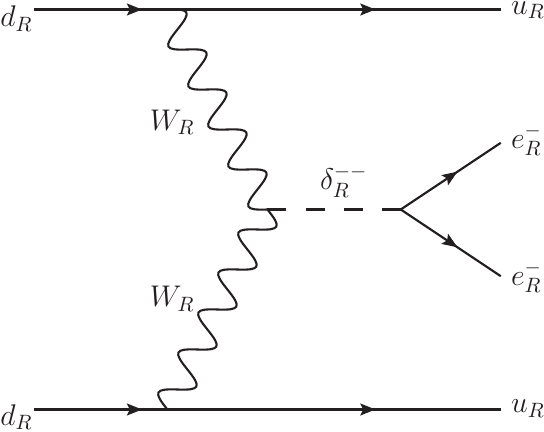}
         \caption{ }
     \end{subfigure}
     \hfill
     \begin{subfigure}[b]{0.22\textwidth}
         \centering
         \includegraphics[width=\textwidth]{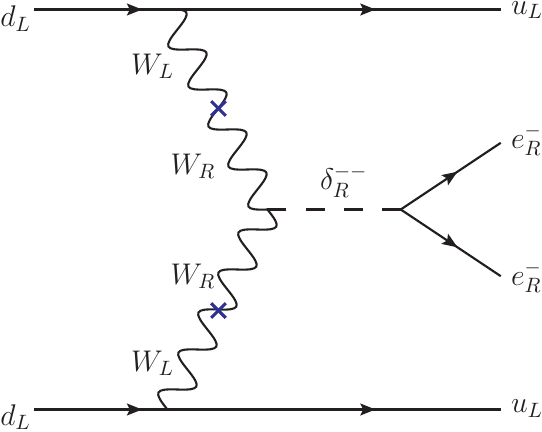}
         \caption{ }
     \end{subfigure}
     \hfill\par\bigskip
     \begin{subfigure}[b]{0.22\textwidth}
         \centering
         \includegraphics[width=\textwidth]{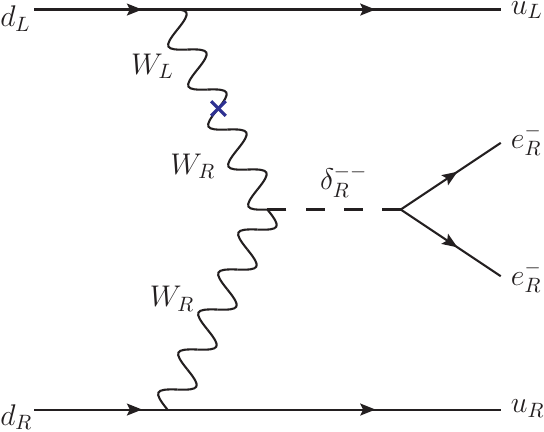}
         \caption{ }
     \end{subfigure}
\caption{Feynman diagrams contributing to $\onbb$ decay in the LRSM. The cross vertices $(\bm\times)$ in between $W_L-W_R$ propagators denote the $W_L-W_R$ mixing.}
\label{fig:feyn-0nbb}
\end{figure}

The Feynman diagrams for the contributions to $\onbb$ decay are shown in panels (a)-(d) of Fig. \ref{fig:feyn-0nbb}. We also include the contributions from right-handed doubly-charged scalar $\delta_R^{--}$ in panels (e)-(g), while those from $\delta_L^{--}$ are omitted. Moreover, contributions arising from mixing between left- and right-handed neutrinos are significantly smaller than those from the right-handed neutrino~\cite{deVries:2022nyh}.\\

The inverse half-life can be expressed as
\begin{align}
\left(T^{0\nu}_{1/2}\right)^{-1} =\ & g_A^4 \Big[ G_{01} \left( |\mathcal A_{L}|^2 + |\mathcal A_{R}|^2 \right)\nn\\
&- 2 (G_{01} - G_{04}) \, \textrm{Re} \{\mathcal A_{L}^* \mathcal A_{R}\} \Big]\;,
\label{eq:Tinv}
\end{align}
where the coupling constant $g_A=1.271$, and $G_{01}$, $G_{04}$ denote the phase-space factors~\cite{Cirigliano:2018yza}. The amplitudes $\mathcal{A}_{L,R}$ are defined as
\begin{align}
    \mathcal{A}_L &= \sum_{i=1}^3 \mathcal{A}_L(m_{i}) \;,\\
    \mathcal{A}_R &= \sum_{i=4}^6 \mathcal{A}_R(m_{i}) + \mathcal{A}_R (M_{\delta_R^{\pm\pm}})\;,
\end{align}
where $m_{i}$ are the active ($i=1,2,3$)  and sterile ($i=4,5,6$) neutrino masses in Eq.~\eqref{eq:mdiagU}. The subamplitudes $\mathcal{A}_{L,R}(m_i)$ are given by~\cite{deVries:2022nyh, Dekens:2020ttz}
\begin{align}
\label{eq:ampL_dim-6}
\mathcal{A}_L(m_{i}) &= -\dfrac{m_{i}}{4m_e}[\mathcal{M}_V(m_{i})+\mathcal{M}_A(m_{i})]\left(C_{{\rm VLL}}^{(6)} \right)_{ei}^2\;,\\
\label{eq:ampR_dim-6}
\mathcal{A}_R(m_{i}) &= -\dfrac{m_{i}}{4m_e}\bigg\{\mathcal{M}_V(m_{i})\left(C_{{\rm VRR}}^{(6)}+C_{{\rm VLR}}^{(6)} \right)_{ei}^2\nn\\
&\qquad\quad+\mathcal{M}_A(m_{i})\left(C_{{\rm VRR}}^{(6)}-C_{{\rm VLR}}^{(6)} \right)_{ei}^2\bigg\}\;.
\end{align}
Here, $m_e$ is the electron mass, and $\mathcal M_{A,V}$ are nuclear matrix elements (NMEs). More details can be found in Refs.~\cite{Prezeau:2003xn, Cirigliano:2017djv, Cirigliano:2018yza, Dekens:2020ttz, deVries:2022nyh}. \\

Notice that the left- and right-handed contributions are proportional to the light-neutrino and the right-handed neutrino masses, respectively. The Wilson coefficients in the mass basis at the scale $\mu=M_W$ are 
\begin{align}
C_{{\rm VLL}}^{(6)}(M_W) &= -2V_{ud} PU \;,\\
C_{{\rm VLR}}^{(6)}(M_W) &= V_{ud} \left(v^2 C_L^{(6)}(M_{W_R}) \right) P_sU^* \;,\\
C_{{\rm VRR}}^{(6)}(M_W) &= \left(v^2 C_R^{(6)}(M_{W_R}) \right) P_sU^* \;.
\end{align}

The Wilson coefficients $C_{L,R}^{(6)}(M_{W_R})$ are obtained by integrating out the $W_R$ boson, and they are
\begin{align}
C_R^{(6)}(M_{W_R}) & = - \dfrac{1}{v_R^2} V_{ud}^R\;,\\
C_L^{(6)}(M_{W_R}) & = 2 \dfrac{\xi }{1+\xi^2} \dfrac{C_R^{(6)}(M_{W_R})}{V_{ud}^R}\;. \label{eq:CL6}
\end{align}

The projectors in the flavor basis $P$ and $P_s$ are
\begin{align}
P = \left( \mathcal{I}_{3\times 3} \quad 0_{3\times 3} \right)\;,\quad
P_s = \left(0_{3\times 3} \quad  \mathcal{I}_{3\times 3}  \right)\;.
\end{align}
On the other hand, the subamplitude
\begin{align}
\mathcal{A}_{R}(M_{\delta_R^{\pm\pm}}) = \dfrac{m_N^2}{m_e v}\mathcal{M}_R^{(9)}
\end{align}
in the EFT approach with dimension-9 effective operators~\cite{Prezeau:2003xn} are explicitly given in Ref.~\cite{Cirigliano:2018yza}. Here, $m_N\sim1\GeV$ is the nucleon mass. For conciseness, we will only repeat the Wilson coefficients of the relevant dimension-9 effective operators~\cite{deVries:2022nyh} as follows:
\begin{align}
C_{1R}^{(9)\prime} (M_{\delta_R^{\pm\pm}}) & = -\dfrac{g_R^4 v^5}{4 M_{W_R}^4} (V_{ud}^R)^2 \dfrac{m_i}{M_{\delta_R^{\pm\pm}}^2} (P_s U^*)^2_{ei}\;, \label{eq:C1R9prime}\\
C_{1R}^{(9)}(M_W) & =  \left( \dfrac{V_{ud}}{V_{ud}^R} \right)^2 \left( \dfrac{\zeta}{\lambda} \right)^2 C_{1R}^{(9)\prime} (M_{\delta_R^{\pm\pm}})\;,\label{eq:C1R9}\\
C_{4R}^{(9)}(M_W) & = 2 \left( \dfrac{V_{ud}}{V_{ud}^R} \right) \dfrac{\zeta}{\lambda}  C_{1R}^{(9)\prime} (M_{\delta_R^{\pm\pm}})\;. \label{eq:C4R9}
\end{align}

The Wilson coefficients $C_{1R}^{(9)\prime} (M_{\delta_R^{\pm\pm}})$ and $C_{1R,4R}^{(9)}(M_W)$ are defined at different scales. The QCD renormalization-group-evolution equations of them and the other Wilson coefficients above are given in Refs.~\cite{Cirigliano:2018yza, deVries:2022nyh}. Note that the contributions of right-handed neutrinos with masses $m_{4,5,6} > \Lambda_\chi$ 
can also be integrated out, rendering dimension-9 effective operators~\cite{Li:2020flq}. The resulting contributions to $\onbb$ decay have also been included in Eq.~\eqref{eq:ampL_dim-6} in the framework developed in Refs.~\cite{deVries:2022nyh, Dekens:2020ttz}.\\

Notice that the CP phase $\alpha$ can contribute to the Wilson coefficients $C_L^{(6)}(M_{W_R})$ and $C_{1R,4R}^{(6)}(M_W)$ as a factor of $e^{-{\rm i}\alpha}$~\cite{deVries:2022nyh}. However, the amplitude for $0\nu\beta\beta$ decay depends on the product of these coefficients with the elements of the PMNS matrix. By marginalizing over the Majorana phases, a non-zero value of $\alpha$ will not affect the results of the $0\nu\beta\beta$ decay rate that we have obtained.

\subsection{Parity-violating Møller scattering}
\label{subsec:moller}
In Møller scattering~\cite{Derman:1979zc}, parity-violating asymmetries in polarized electron scattering can constrain the 4-electron effective vertex in a model-independent way. For that purpose, the MOLLER Collaboration proposes to measure the 
asymmetry in the scattering of polarized electrons off unpolarized electrons using the $11\GeV$ beam in Hall A at Jefferson Lab to measure the electron's weak charge $Q_W^e$ to an overall accuracy of 2.4\%~\cite{MOLLER:2014iki}. 
This measurement could represent an enhancement in fractional precision exceeding five times that of the only other similar one done by the E158 experiment at SLAC \cite{SLACE158:2005uay}.
The MOLLER sensitivity can be expressed as the following lower bound:
\begin{align}
\frac{\Lambda}{\sqrt{|g_{RR}^2-g_{LL}^2|}} ~\simeq~ 7.5\TeV \,.
\label{eq:4fermi}
\end{align}
Here, $g_{LL,RR}$ are the coupling constants involving left-handed and right-handed electrons in the 4-fermion effective interaction defined below
\begin{equation}
\mathcal{L}_{\mathit{eff}}=\sum_{i,j=L,R} \frac{g_{ij}^{2}}{2\Lambda^2} (\bar{e}_{i} \gamma^{\mu} e_i)(\bar{e}_j \gamma_{\mu} e_j)\,,
\label{eq:L_4fermion}
\end{equation}
where $\Lambda$ is the mass scale of new physics~\cite{MOLLER:2014iki, Eichten:1983hw}. \\

\begin{figure}[t]
\centering
\includegraphics[width=0.55\columnwidth]{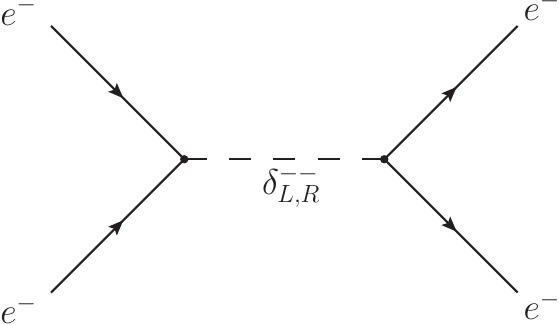}
\caption{Doubly-charged scalar contributions to parity-violating Møller scattering. Notice that each vertex violates the lepton number by two units.}
\label{fig:feyn-moller}
\end{figure}

In the context of the LRSM, one can relate parity-violating Møller scattering, which conserves the lepton number, to $\onbb$ decay that violates the lepton number by two units. 
The leptonic Yukawa interactions in Eq.~\eqref{eq:Lag-Yukawa} allow for both the left- and the right-handed doubly-charged scalars to contribute to 
Møller scattering. These contributions consist of an $s$-channel exchange depicted in Fig.~\ref{fig:feyn-moller}.
The corresponding parity-violating effective Lagrangian can be written as~\cite{Dev:2018sel}
\begin{align}
{\cal L}_{\rm PV}^{\mathit{eff}} = \frac{|(f_{L})_{ee}|^2}{M_{\delta_L^{\pm\pm}}^2}(\bar{e}_L\gamma^\mu e_L)(\bar{e}_L\gamma_\mu e_L) ~+~ (L \leftrightarrow R) \,,
\label{eq:L_PV}
\end{align}
where we have used the Firez identity~\cite{Nieves:2003in,Liao:2012uj}
\begin{align}
    \left(\bar e_L e_L^c\right)\left(\bar e_L e_L^c\right) =\dfrac{1}{2} \left(\bar e_L \gamma^\mu e_L\right)\left(\bar e_L \gamma_\mu e_L\right)\,.
\end{align}

It is worth noting that there exists an additional contribution to
Møller scattering from the $t$-channel exchange of the neutral scalar from the scalar bidoublet. However, the corresponding amplitude does not interfere with the SM amplitude and thus contributes at the $1/\Lambda^4$ order, which can be easily verified with the following Fierz identity
\begin{align}
    \left(\bar e_L e_R\right)\left(\bar e_L e_R\right) =-\dfrac{1}{2} \left(\bar e_L \gamma^\mu e_L\right)\left(\bar e_R \gamma_\mu e_R\right)\,.
\end{align}

Given the severe constraints on the mass of the neutral scalar from the scalar bidoublet~\cite{Bertolini:2019out,Dekens:2021bro}, their contributions to the Møller scattering can be safely neglected.\\

In principle, the 
effective Lagrangian in Eq.~\eqref{eq:L_PV} has left- and right-handed contributions. However, the severe constraints in Eq.~\eqref{eq:cLFV-constraints} from the searches for $\mu \to eee$ and $\mu \to e\gamma$ imply that both the couplings $(f_L)_{ee}$ and $(f_L)_{e\mu}$ are tiny or $\delta^{\pm\pm}_L$ is heavy. Therefore, only the contribution from $\delta_R^{\pm\pm}$ will be considered in light of the MOLLER prospect. \\

We can directly match the corresponding expressions in Eq.~\eqref{eq:L_PV} 
 with Eq.~\eqref{eq:4fermi}, where the effective mass scale is $\Lambda=M_{\delta_R^{\pm\pm}}$ and the effective couplings are $|g_{RR}|^2=2|\fRee|^2$ and $g_{LL}=0$. Therefore, the parity-violating precision measurement would constrain the ratio between the right-handed doubly-charged scalar mass $M_{\delta_R^{\pm\pm}}$ and its Yukawa coupling to electrons $(f_R)_{ee}$~\cite{MOLLER:2014iki},
 \begin{align}
\frac{M_{\delta_R^{\pm\pm}}}{|(f_R)_{ee}|} ~\gtrsim~ 7.6\TeV
\label{eq:moller_limit}
\end{align}
at $95\%$ confidence level (CL)\,\footnote{Notice that Ref.~\cite{Dev:2018sel} misses a factor of two in the mapping into the 4-fermion effective interaction in Eq. \eqref{eq:L_4fermion} that is corrected here.}.

\section{Results and Discussion }
\label{sec:res_disc}

This section discusses in detail the results of $\onbb$ decay in the EFT approach and its interplay with the other searches and measurements. Notice that the right-handed neutrinos could have masses of the order of the right-handed scale $v_R$, but can also be significantly lower.
To establish a concrete $f_R$-correlation between the observables under consideration, we will focus our study on the effects of the right-handed neutrino $\nu_4$ with the mass
\begin{align}
M_{N_R}\equiv m_4=\sqrt{2}\,v_R\,|(f_R)_{ee}|\,,
\label{eq:MNmass}
\end{align}
while the other right-handed neutrinos decouple. 
On the other hand, from Eq.~\eqref{eq:mneutrinos}, the masses of the active neutrinos are completely determined once the lightest active neutrino mass is chosen and the hierarchy is specified. The oscillation parameters of active neutrinos encoded in the PMNS matrix are quoted from Ref. \cite{deSalas:2020pgw}.\\

We will start by investigating the sensitivity of $\onbb$-decay searches.
We adopt the central values of the oscillation parameters in Table 3 from Ref. \cite{deSalas:2020pgw} and marginalize over the Majorana phases $\lambda_{1,2}$ 
in calculating the $\onbb$-decay rate. The lightest active neutrino mass is randomly generated with the upper bound $m_{\nu_{\min}}<0.05\eV$.
Hence, the inverse half-life in Eq.~\eqref{eq:Tinv} depends only on a few parameters 
\begin{itemize}
\item The mass of the doubly-charged scalar $M_{\delta_{R}^{\pm\pm}}$ and its Yukawa coupling to electrons $|(f_R)_{ee}|$.
\item The mass of right-handed gauge boson $M_{W_R}$ and $\zeta$, the $W_L-W_R$ mixing angle\,\footnote{Alternatively, the ratio of the vevs $\xi=\tan\beta$ can also be used, instead of the mixing angle $\zeta$, in virtue of Eq.~\eqref{eq:tan_zeta__lambda_sin2beta}.}.
\end{itemize}

Notice that these parameters completely determine the right-handed neutrino mass $M_{N_R}$ by combining Eqs.~\eqref{eq:MWmass} and~\eqref{eq:MNmass}, $M_{N_R}=(2/g_R)\,M_{W_R}\,|(f_R)_{ee}|$. We do not restrict ourselves to a specific underlying model but adopt the assumption $g_R =  2g_L/3$ inspired by Ref.~\cite{Deppisch:2014zta} in the numerical analysis for illustrative purposes. Results for other scenarios, such as $g_R = g_L$, can be easily derived within our framework.
Since parity-violating Møller scattering is only sensitive to the subset of parameters $\{\MDelta,|\fRee|\}$ (cf. Eq.~\eqref{eq:moller_limit} and subsequent discussion), we choose different benchmark scenarios for $\MWR$ and the left-right mixing angle $\zeta$. These values were selected to be consistent with the constraints discussed in Sec.~\ref{sec:exper_constr_low-E}.\\

\begin{figure*}[!htb]
\centering
\begin{subfigure}[!htb]{\textwidth}
\includegraphics[height=0.32\textwidth]{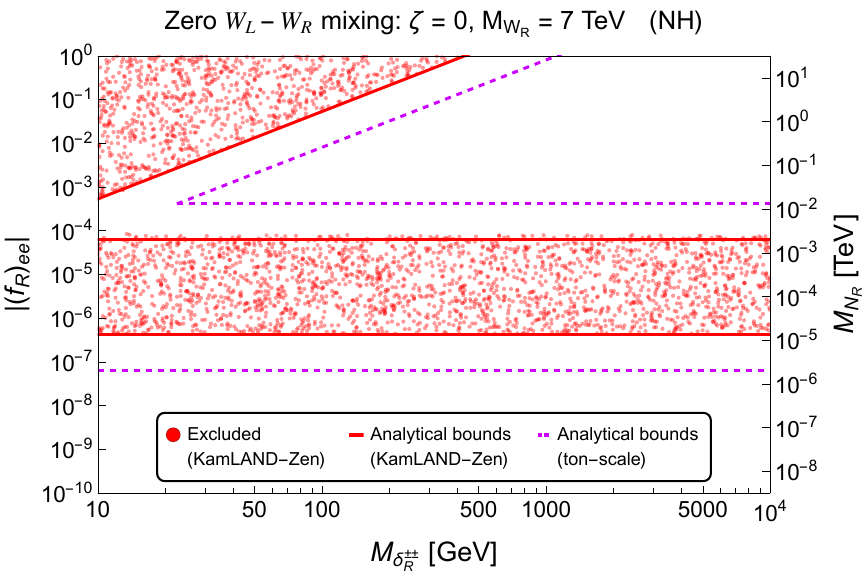}\hspace{1ex}
\includegraphics[height=0.32\textwidth]{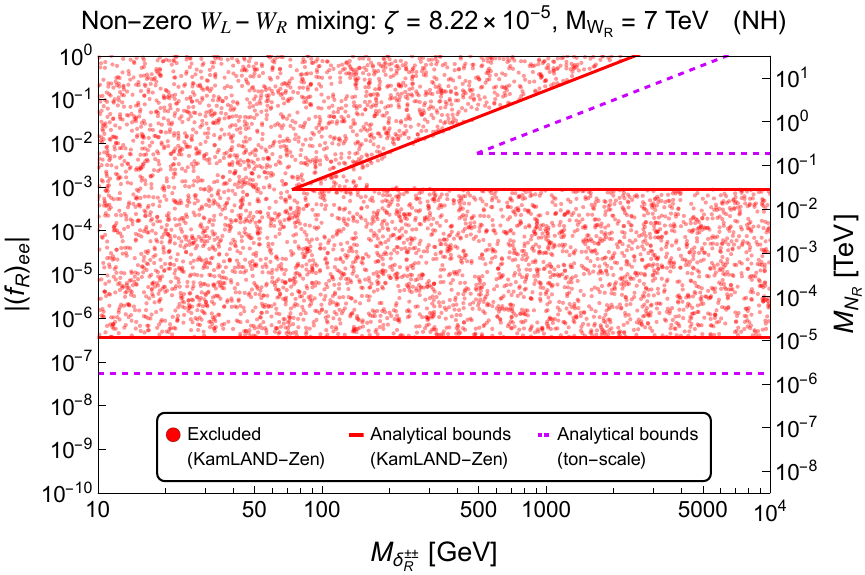}
\vspace{1em}
\end{subfigure}
\begin{subfigure}[!htb]{\textwidth}
\centering
\includegraphics[height=0.32\textwidth]{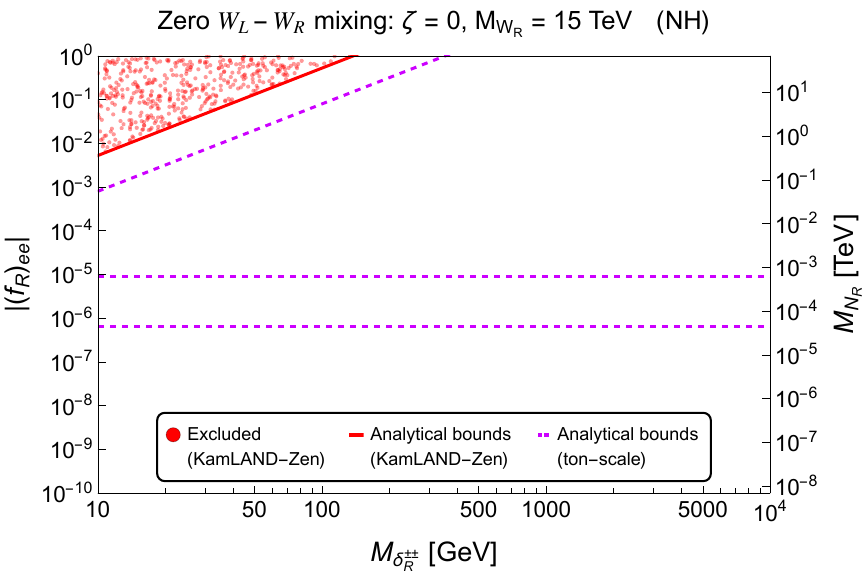}\hspace{1ex}
\includegraphics[height=0.32\textwidth]{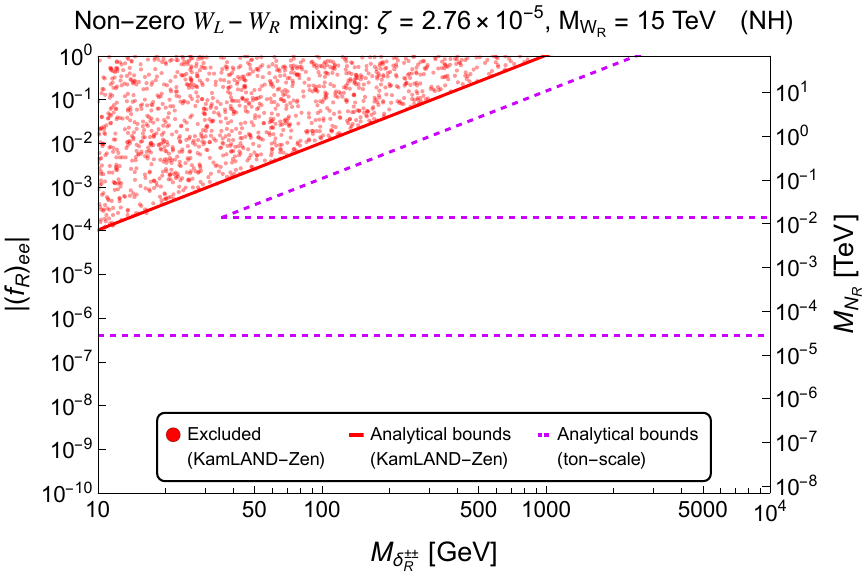}
\end{subfigure}
\caption{
Parameter scan of $\onbb$-decay constraints on the right-handed doubly-charged scalar mass $M_{\delta_R^{\pm\pm}}$ and the Yukawa coupling $|(f_R)_{ee}|$ in the LRSM with zero (left column) and non-zero (right column) $W_L-W_R$ mixing. 
The active neutrino masses are assumed in the normal hierarchy (NH), and the right-handed $W$ boson mass is chosen as $M_{W_R}=7\TeV$ (top row) and $\MWR=15\TeV$ (bottom row).
We take $\xi=0$ and $0.35$ for the left- and right-handed plots, respectively, and use Eq.~\eqref{eq:tan_zeta__lambda_sin2beta} to characterize the results in terms of the physical mixing angle $\zeta$. The right-handed neutrino mass $M_{N_R}$ is also displayed as a secondary vertical axis. The red points (regions above or between the solid red lines) are excluded by the most recent KamLAND-Zen results~\cite{KamLAND-Zen:2022tow}. The solid red lines correspond to the exclusion limits calculated using an analytical estimation (see text for discussion). The dashed magenta lines correspond to the same exclusion limits using the prospect by ton-scale experiments~\cite{Agostini:2017jim}.}
\label{fig:0nubb_MWRall}
\end{figure*}

Using the expressions in Sec.~\ref{subsec:0nbb}, we obtain constraints from the $\onbb$-decay searches in Fig.~\ref{fig:0nubb_MWRall} for the active neutrino masses in the normal hierarchy (NH)\,\footnote{There are no significant differences when the inverted hierarchy is chosen, so we omit the results for the sake of brevity.}. A parameter scan over $\MDelta$ and $|\fRee|$ is performed to predict $T_{1/2}^{0\nu}$ and compare it with the most recent KamLAND-Zen results ($T_{1/2}^{0\nu}>2.3\times 10^{26}$ years at 90\% C.L.~\cite{KamLAND-Zen:2022tow}) to obtain the excluded parameter regions shown as red points. The solid red lines bound these regions, and they correspond to theoretical estimates of the $\onbb$ amplitude based on the different contributions (doubly-charged scalar, active neutrino, and sterile neutrino) in Eq.~\eqref{eq:Tinv}, as we will explain in the next paragraphs. We use the same theoretical estimate to draw the exclusion boundary, using ton-scale prospects that are expected to be sensitive to a half-life of $10^{28}$ years~\cite{Agostini:2017jim}; these results are shown as dashed magenta lines.\\

The top row in Fig.~\ref{fig:0nubb_MWRall} shows two separate regions that are consistent with the observed $\onbb$-decay constraints. The top region accommodates large Yukawa couplings and heavy doubly-charged scalars, while the lower region is independent of the doubly-charged scalar mass and is consistent with small couplings. The gap between these two regions corresponds to the transition between a heavy and a light right-handed neutrino description in the amplitude in Eq.~\eqref{eq:ampR_dim-6}. To better understand these results, we will analyze the different contributions to the amplitude and identify the ranges where the individual contributions shown in Fig.~\ref{fig:feyn-0nbb} dominate.\\

The specific dependence of $T_{1/2}^{0\nu}$ as a function of $|\fRee|$ can be determined by analyzing the different contributions to the amplitude in Eq.~\eqref{eq:ampR_dim-6}. There are four regimes to be considered:
\begin{itemize}
\item \textbf{Doubly-charged scalar regime.} For a large Yukawa coupling, the doubly-charged scalar contribution to the amplitude dominates the inverse half-life in Eq.~\eqref{eq:Tinv}. By analyzing diagrams (e), (f), and (g) in Fig.~\ref{fig:feyn-0nbb}, we notice that the inverse half-life resulting after integrating out $W_R$ and $\delta_R^{\pm\pm}$ will be proportional to
\begin{align}
(T_{1/2}^{0\nu})^{-1} &\propto \dfrac{1}{\MDelta^4\,\MWR^6}\,|\fRee|^2\,,
\label{eq:InvHalfLife_MDelta}
\end{align}

where  the second relation in Eq.~\eqref{eq:MWmass} has been used. The overall factor is given by the product of the phase-space factor, NMEs, and low-energy constants. It also depends on the $W_L-W_R$ mixing and indicates that in this regime, for a larger $\MDelta$, a larger $|\fRee|$ is allowed by the constraints of $\onbb$-decay experiments, shown as the diagonal boundaries in Fig.~\ref{fig:0nubb_MWRall}. It is worth noting that this region of the parameter space aligns with the findings in Ref.~\cite{Dev:2018sel}, which focused only on the effects of the doubly-charged scalar as new physics contributions to $\onbb$-decay. Additionally, when the left-right mixing is turned on, the contribution to the inverse half-life increases so that the region is shrunk to the right.

\item \textbf{Heavy right-handed neutrino regime.} For a moderate Yukawa coupling, the contribution from right-handed neutrinos dominates the $\onbb$-decay amplitude, and the right-handed neutrino mass is large enough to be integrated out. From diagrams (b), (c), and (d) in Fig.~\ref{fig:feyn-0nbb}, we estimate the inverse half-life of $\onbb$ decay as 
\begin{align}
(T_{1/2}^{0\nu})^{-1} 
&\propto 
\dfrac{1}{\MWR^{10}}
\,\dfrac{1}{|\fRee|^2}\,,
\label{eq:InvHalfLife_MNheavy}
\end{align}
where we have used the relation in Eq.~\eqref{eq:MNmass} to obtain the extra power of $\MWR$ in the amplitude's denominator. This contribution is independent of $\MDelta$ and denoted by the upper horizontal boundary in Fig.~\ref{fig:0nubb_MWRall}. The contribution to the inverse half-life also increases if the left-right mixing is turned on so that a larger $|\fRee|$ is preferred by comparing the left and right panels in Fig.~\ref{fig:0nubb_MWRall}.

\item \textbf{Light right-handed neutrino regime.} The small Yukawa coupling region corresponds to the scenario where the contribution from right-handed neutrinos still dominates the $\onbb$-decay amplitude. However, the right-handed neutrino mass is small enough so that the light right-handed neutrino is not integrated out. By analyzing diagrams (b), (c), and (d) in Fig.~\ref{fig:feyn-0nbb}, we can estimate the inverse half-life of $\onbb$ decay as
\begin{align}
(T_{1/2}^{0\nu})^{-1} 
&\propto \dfrac{M_{N_R}^2}{\langle p\rangle^4  \MWR^{8}} = \dfrac{|\fRee|^2}{\langle p\rangle^4  \MWR^{6}}\,,
\label{eq:InvHalfLife_MNlight}
\end{align}
where $\langle p\rangle$ denotes the averaged momentum transfer, which is about $100-200\MeV$.

\item \textbf{Left-handed neutrino regime.} The vanishingly small Yukawa coupling region corresponds to the scenario where the dominant contribution to $\onbb$ decay comes from the standard mechanism. By analyzing diagram (a) in Fig.~\ref{fig:feyn-0nbb}, we can estimate the inverse half-life as
\begin{align}
(T_{1/2}^{0\nu})^{-1} 
&\propto \frac{G_F^4\,m_\nu^2}{\langle p\rangle^4}\,,
\label{eq:InvHalfLife_Mnu}
\end{align}
where $\langle p\rangle$ denotes the averaged momentum transfer, which is about $100-200\MeV$, $G_F\sim M_{W_L}^{-2}$ is the Fermi constant, and $m_{\nu}$ is neutrino mass.
\end{itemize}

\begin{figure*}[!htb]
  \centering
  \includegraphics[width=0.45\textwidth]{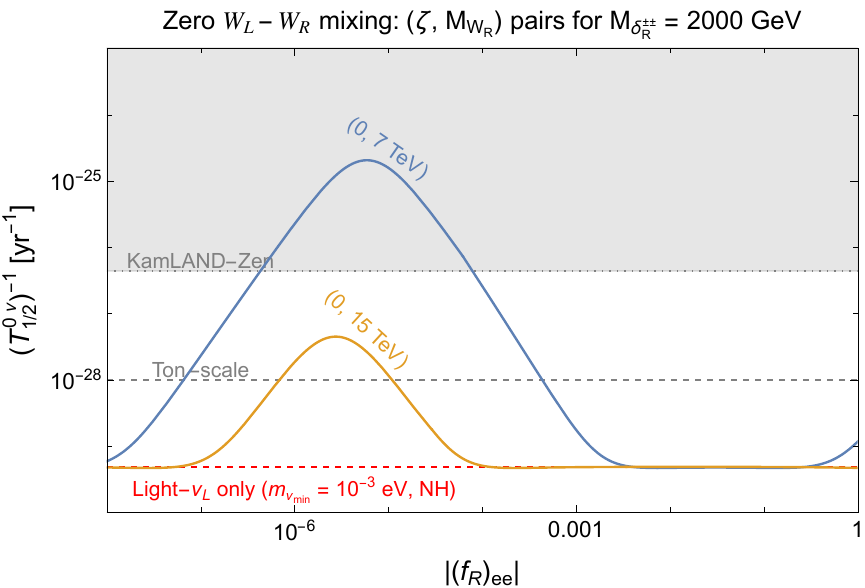}\hspace{1ex}
  \includegraphics[width=0.45\textwidth]{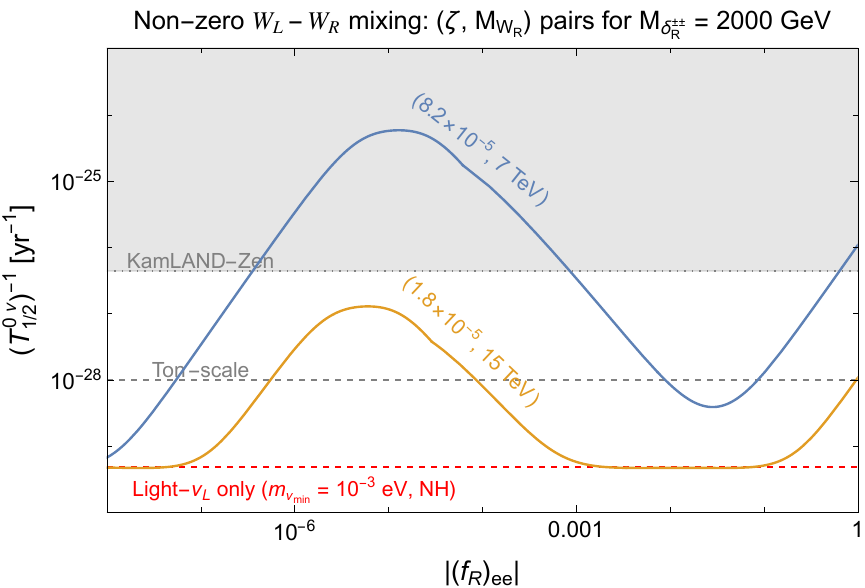}
  \caption{Inverse 
  half-life $(T^{0\nu}_{1/2})^{-1}$ as a function of the Yukawa coupling $|\fRee|$ for zero (left) and non-zero $W_L-W_R$ mixing (right). We fix $\xi=0$ and $\xi=0.35$ for the left and right plots, respectively. We also fix the mass of the doubly-charged scalar at $\MDelta=2000\GeV$. The two curves in the figure show the results for $\MWR=7\TeV$ (blue) and $\MWR=15\TeV$ (yellow). We also show the corresponding values of the mixing angle $\zeta$. The inverse half-life considering only the light left-handed neutrino contributions is shown in red. The latest KamLAND-Zen exclusion limits are represented by the grey shaded area, with the future ton-scale experiment bounds indicated for comparison.}
  \label{fig:T1/2_fRee}
\end{figure*}

\begin{figure*}[!htb]
\centering
\includegraphics[width=0.45\textwidth]{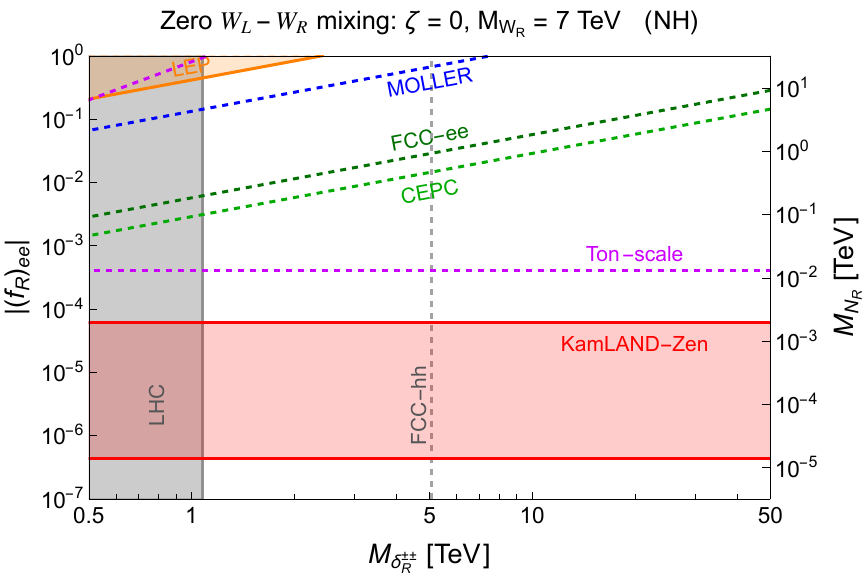}\hspace{1ex}
\includegraphics[width=0.45\textwidth]{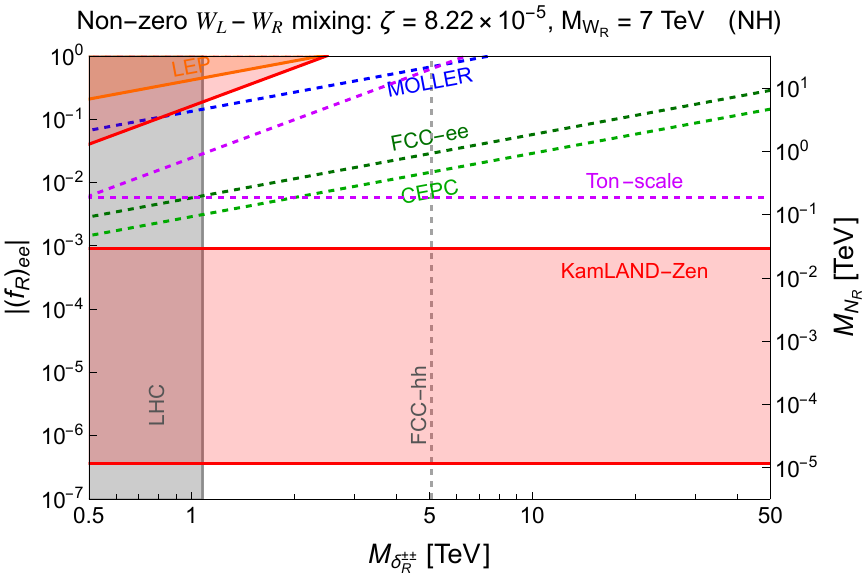}
\caption{Combined sensitivities to the right-handed doubly-charged scalar mass $M_{\delta_R^{\pm\pm}}$ and the Yukawa coupling $|(f_R)_{ee}|$ in the
$\onbb$-decay, MOLLER, and collider experiments in the LRSM. The normal hierarchy is assumed. We fix the mass of the $W_R$ boson to $M_{W_R}=7\TeV$ and present scenarios with zero (left) and non-zero (right) $W_L-W_R$ mixing. The right-handed neutrino mass $M_{N_R}$ is also displayed as a secondary vertical axis. The MOLLER prospect~\cite{MOLLER:2014iki} is shown as a dashed blue line, and $\onbb$ decay limits from current and future experiments are shown as solid red and dashed magenta lines, respectively. The figure also displays the direct searches at the LHC (solid gray), pair production prospect at the FCC-hh (dashed gray), and Bhabha scattering limits at LEP (solid orange), the CEPC (dashed green), and the FCC-ee (dashed dark green) prospects. 
}
\label{fig:Moller&0nbb_MWR7TeV}
\end{figure*}

To understand the gap in the top row in Fig. \ref{fig:0nubb_MWRall}, let us consider $\invhalf$ as a function of $|\fRee|$ for $\MDelta = 2\TeV$, $\MWR=7\TeV$, and zero/non-zero $W_L-W_R$ mixing, which corresponds to taking vertical slices in both panels of the top row in Fig. \ref{fig:0nubb_MWRall}. \\

We present these results in Fig.~\ref{fig:T1/2_fRee} as a solid blue line. For the left and right panels, we consider $\xi=0$ and 0.35, respectively, and use Eq.~\eqref{eq:tan_zeta__lambda_sin2beta} to characterize the results in terms of the physical mixing angle $\zeta$ and $\MWR$ in the case of a non-zero $W_L-W_R$ mixing.
To show $\invhalf$ in the standard mechanism, a red dashed curve is added. The grey shaded area represents the latest KamLAND-Zen exclusion limit, with the future ton-scale experiment bound indicated by the dashed gray line for comparison. \\

From Fig.~\ref{fig:T1/2_fRee}, we find that in the large coupling region, where the amplitude is dominated by the doubly-charged scalar contribution, $\invhalf$ decreases as $|\fRee|$ is reduced, remaining consistent with current experimental bounds. We then observe a transition to a second regime, where the heavy right-handed neutrino contribution dominates the amplitude. In this regime, $\invhalf$ increases as $|\fRee|$ decreases. Consequently, for values of $|\fRee| \lesssim 10^{-3}$, the model is excluded by current experimental data. However, around $|\fRee| \sim 10^{-5}$, the right-handed neutrino becomes sufficiently light to alter the functional dependence of 
$\invhalf$, causing it to decrease again as $|\fRee|$ is further reduced. As expected, for values of $|\fRee| \lesssim 10^{-7}$, $\invhalf$ becomes compatible with current experimental bounds again\,\footnote{
For such a small Yukawa coupling $(f_R)_{ee}$, the mass of the right-handed neutrino that couples to the electron is $M_{N_R} \gtrsim \mathcal{O}({\rm MeV})$. One might be concerned about whether the constraints from various experimental searches~\cite{Bolton:2019pcu,Alves:2023znq} can be evaded. In the type-I seesaw mechanism, the active-sterile mixing is of the order $M_D/M_R \sim \sqrt{M_\nu/M_R}$, where $M_R = \sqrt{2}v_R {\bf f_R}$. Consequently, this mixing can vary significantly depending on the flavor structure of ${\bf f_R}$. 
On the other hand, a small region of parameter space for $140\MeV \lesssim M_{N_R}\lesssim 500\MeV$ is excluded by recent analyses of meson decays~\cite{Alves:2023znq}, particularly for $M_{W_R}= 7\TeV$. However, this exclusion is unlikely to have a significant impact on our results.
}.\\

In Fig. \ref{fig:T1/2_fRee}, we also include the curves of $\invhalf$ as a function of $|\fRee|$ for $\MWR=15\TeV$ in yellow. As we increase the value of $\MWR$, the BSM effects are suppressed, $\invhalf$ is smaller, and gets closer to the prediction in the standard mechanism. In particular, for $\MWR = 15\TeV$, the whole parameter region of $|\fRee|$ satisfies the KamLAND-Zen constraint. Turning into the effects of the mixing angle, as shown in Eqs.~\eqref{eq:CL6}, \eqref{eq:C1R9prime}--\eqref{eq:C4R9}, $\zeta$ impacts the sub-amplitudes involving a right-handed neutrino exchange. In simple terms, compared with the $\zeta=0$ case, a non-zero mixing angle extends the parameter region where the light right-handed neutrino description is applicable and reduces the parameter region where the heavy right-handed neutrino description is relevant. By comparing the left and right panels in Fig. \ref{fig:T1/2_fRee}, we notice that the peaks of $\invhalf$ are higher and further to the right for the cases of $\zeta\neq0$, respectively, whereas the local minima are higher and closer to the peaks for the cases of $\zeta\neq0$, respectively. We can now analyze the entire Fig. \ref{fig:0nubb_MWRall} in light of these findings. As expected, we noticed that the gap between the allowed regions narrows with increasing $W_R$ boson mass (from left to right), widens with non-zero $W_L-W_R$ mixing (from top to bottom), and disappears completely for $\MWR=15\TeV$ for the current $\onbb$ decay constraints.\\

\begin{figure*}[t!]
  \centering
  \includegraphics[width=0.45\textwidth]{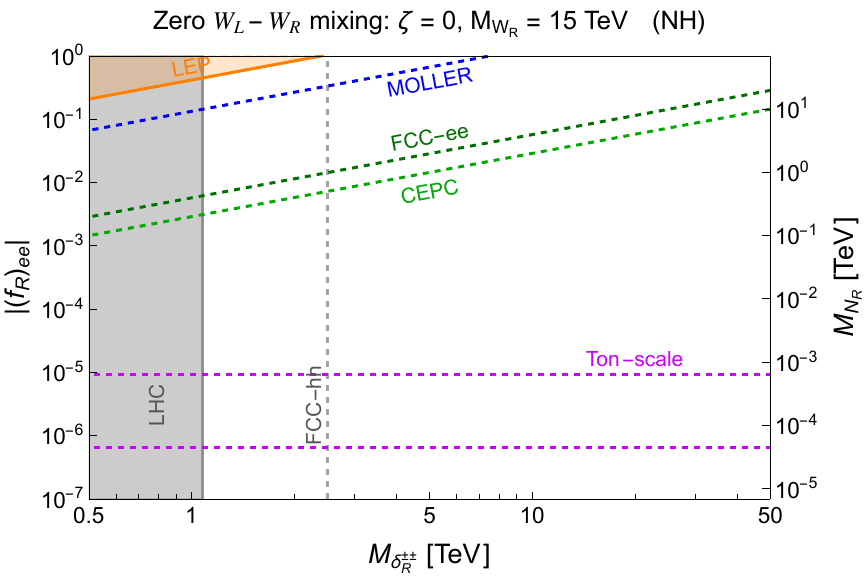}\hspace{1ex}
  \includegraphics[width=0.45\textwidth]{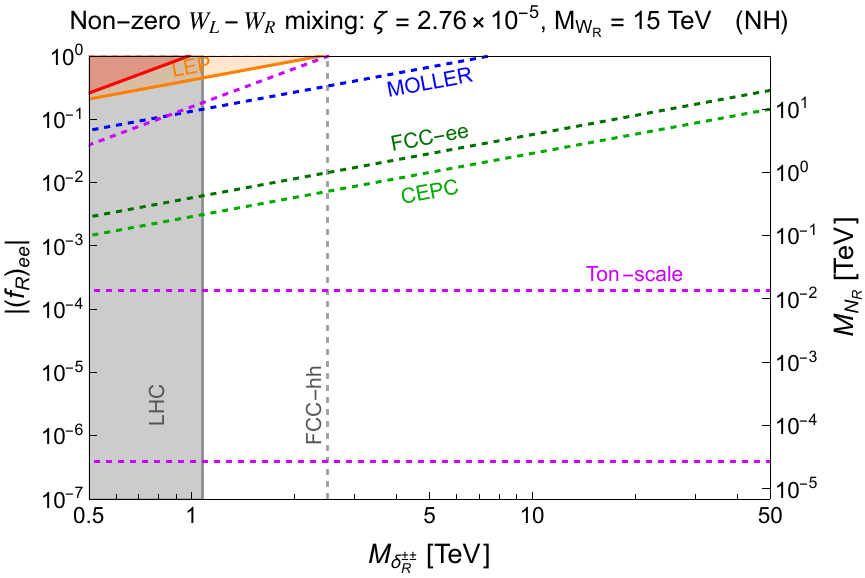}
  \caption{Same as in Fig.~\ref{fig:Moller&0nbb_MWR7TeV}, combined constraints from LNC and $\onbb$-decay experiments on the right-handed doubly-charged scalar mass $M_{\delta_R^{\pm\pm}}$ and the Yukawa coupling $|(f_R)_{ee}|$ in the parity-violating LRSM in the normal hierarchy with $M_{W_R}=15\TeV$ for zero (left) and non-zero (right) $W_L-W_R$ mixing.}
  \label{fig:Moller&0nbb_MWR15TeV}
\end{figure*}

To illustrate the complementary sensitivities of $\onbb$-decay searches and LNC processes in both low-energy experiments and high-energy colliders, we present results based on current constraints and future prospects, as described in Sec. \ref{sec:exper_constr_high-E} and Sec. \ref{sec:exper_constr_low-E}. The experimental sensitivities to $\MDelta$ and $|\fRee|$ are shown in Figs.~\ref{fig:Moller&0nbb_MWR7TeV} and~\ref{fig:Moller&0nbb_MWR15TeV} for zero and non-zero mixing angles. In each plot, the vertical axis on the right gives the values of $M_{N_R}$ that corresponds to $|(f_R)_{ee}|$ using Eq.~\eqref{eq:MNmass}.\\

In these two figures, the $\onbb$ decay exclusion limits 
are shown as red solid lines and correspond to the doubly-charged scalar (top diagonal), heavy right-handed neutrino (upper horizontal), and light right-handed neutrino (lower horizontal) contributions to the amplitude. They were obtained using the analytical expressions in Eqs.~\eqref{eq:InvHalfLife_MDelta}--\eqref{eq:InvHalfLife_MNlight}. The shaded red regions limited by those lines are excluded by the KamLAND-Zen experiment. Similar prospects for ton-scale experiments are included in dashed magenta lines. 
For comparison purposes, we can also calculate the $\onbb$ decay exclusion limit derived in Ref.~\cite{Dev:2018sel} with the left-right mixing angle $\zeta=0$. Rescaling their limit to accommodate the latest KamLAND-Zen results and modifying the corresponding values of $M_{W_R}$\,\footnote{Ref.~\cite{Dev:2018sel} assumed a fixed value of $v_R = 5\sqrt{2}\TeV$. Under the assumption of $g_R=2g_L/3$, this corresponds to $\MWR=2.2\TeV$, which has been excluded by the LHC  searches at Run-2 phase~\cite{ATLAS:2023cjo, CMS:2022yjm, CMS:2021dzb}.} one obtains $|\fRee|/\MDelta^2\lesssim 0.1\times(\MWR/{\rm TeV})^3$, not visible in the range of Fig.~\ref{fig:Moller&0nbb_MWR7TeV}. Notice how the use of an EFT description provides a stronger constraint in the $\onbb$-decay sensitivity compared to the existing literature~\cite{Dev:2018sel}.\\

The MOLLER exclusion limit is shown as a dashed blue line. Its relative sensitivity compared with $\onbb$ decay is influenced by the presence of a $W_L-W_R$ mixing. When the mixing is zero ($\zeta=0$, left panels), the doubly-charged scalar contribution to $\onbb$ decay yields a weaker constraint to both the KamLAND-Zen and ton-scale prospective experiments compared to the MOLLER experiment. Conversely, for non-zero mixing ($\zeta\neq0$, right panels), the MOLLER experiment achieves a stronger (weaker) sensitivity compared to the next generation of $\onbb$ experiments for larger (smaller) $\MDelta$. For example, as seen in Fig.~\ref{fig:Moller&0nbb_MWR7TeV}, a positive result in the MOLLER experiment with a corresponding observation in $\onbb$-decay experiments for $\MDelta\lesssim5\TeV$ would immediately imply a non-zero $W_L-W_R$ mixing value (right panel) over zero mixing (left panel). On the contrary, a positive event in ton-scale $\onbb$ decay experiments without a corresponding signal in the MOLLER experiment for $\MDelta\lesssim5\TeV$ is consistent with both zero and non-zero mixing scenarios. Notice that even since the $\onbb$ decay sensitivity is highly dependent on $\MWR$, the MOLLER experiment can never prove the smaller $|\fRee|$ region associated with the right-handed neutrino-dominated region for $\onbb$ decay.\\

We also include collider searches in Figs.~\ref{fig:Moller&0nbb_MWR7TeV} and~\ref{fig:Moller&0nbb_MWR15TeV}. The current indirect LEP exclusion limit for Bhabha scattering is shown as a solid orange line, with its corresponding exclusion region shaded in orange. The prospects for Bhabha scattering at the CEPC and the FCC-ee are shown as dashed dark green and light green lines, respectively, using the rescaling factor in Eq.~\eqref{eq:factor}. Notice that the difference in luminosities between the two accelerators makes the CEPC bound stronger. These collider searches will reach into the ``funnel'' portion of the parameter space that the $\onbb$ decay cannot access with current envisioned experiments. Direct searches for right-handed doubly-charged scalar $\delta_R^{\pm\pm}$ at hadron colliders are also shown for both current (LHC, grey solid line) and future (FCC-hh, grey dashed line) experiments, as discussed in Sec.~\ref{sec:exper_constr_high-E}. They are sensitive to $\MDelta \lesssim5\TeV $ irrespective of the Yukawa coupling $|\fRee|$. If we conceive a scenario where a positive event in ton-scale $\onbb$ decay experiments is seen but both future leptons and hadrons colliders do not get a corresponding signal, then we can pin down the underlying $\onbb$-decay mechanism to be dominated by the right-handed neutrino exchange and the corresponding parameter region would be a horizontal fringe of smaller $|\fRee|$ in Figs.~\ref{fig:Moller&0nbb_MWR7TeV} and~\ref{fig:Moller&0nbb_MWR15TeV}. \\

\section{Conclusion}
\label{sec:conclusions}

In this work, we have studied the interplay of $\onbb$-decay,  parity-violating Møller scattering, and collider searches in probing the Yukawa interaction of the right-handed doubly-charged scalar to electrons
in the left-right symmetric model. We investigate the sensitivities of the low-energy experiments and the high-energy colliders and find that they complement each other in exploring the region of Yukawa coupling down to $10^{-7}$.\\

We employ the effective field theory framework to calculate the $\onbb$-decay rate with the contributions arising from the mediation of right-handed neutrinos and the right-handed doubly-charged scalar, considering possible non-zero $W_L-W_R$ mixing. Compared to the earlier work of Ref. \cite{Dev:2018sel}, the effective field theory framework provides a more realistic assessment of the $\onbb$-decay sensitivity and yields generally significantly greater parameter space reach.\\

Four regimes are considered depending on the magnitude of Yukawa coupling, or equivalently, the mass of the right-handed neutrino given Eq.~\eqref{eq:MNmass}.
The sensitivities of the KamLAND-Zen and ton-scale experiments are compared with those for the precision measurements of parity violation in the MOLLER experiment.\\

We find that in the absence of $W_L-W_R$ mixing, the MOLLER sensitivity is consistently stronger than that in future ton-scale $\onbb$ decay experiments for the regime where the doubly-charged scalar dominates the $\onbb$ amplitude. For non-zero mixing, however, the relative sensitivity between $\onbb$ decay searches in ton-scale experiments and precision measurements of parity violation in the MOLLER experiments depends on the mass of the doubly-charge scalar. Moreover, since MOLLER cannot prove the small $|\fRee|$ region, $\onbb$ decay experiments have the unique sensitivity to exclude values of $|\fRee|$ as small as $10^{-7}$, independent of the doubly-charged scalar contribution.\\

Finally, the interplay of collider and low-energy searches provides a manner to explore regions that are inaccessible to $\onbb$ decay experiments, as they are currently conceived. Even in the absence of positive results in the next generations of future leptons or hadrons colliders, we are still able to identify the underlying dominant mechanism in $\onbb$ decay and explore its corresponding parameter region.\\

\section*{Acknowledgments}
GL is supported by the National Natural Science Foundation of China under Grant No. 12347105, the Guangdong Basic and Applied Basic Research Foundation (2024A1515012668), 
and SYSU startup funding. MJRM and SUQ were supported in part under the US Department of Energy contract DE-SC0011095. SUQ also acknowledges the support of the U.S. DOE
under Grant No. DE-FG02-00ER41132. SUQ would like to thank the High Energy Theory Group at William \& Mary for their hospitality, where part of this work was conducted.

\appendix

\bibliographystyle{apsrev4-1}
\balance
\biboptions{sort&compress}
\bibliography{Moller_0nbb.bib}

\end{document}